\documentclass{article}
\usepackage{a4wide, cite, amsmath, graphics, subfigure}

\title{Glassy behaviour in a simple topological model}
\author{Lexie Davison\thanks{ldavison@thphys.ox.ac.uk} \hspace{4pt}and David Sherrington\thanks{d.sherrington1@physics.ox.ac.uk}\\Dept. of Physics,
Oxford University\\ \textit{Theoretical Physics, 1 Keble Road, Oxford OX1 3NP,
England}}

\date{\today}
\begin{document}
\maketitle

\begin{abstract}
In this article we study a simple, purely topological, cellular model
which is allowed to evolve through a Glauber-Kawasaki process. We find a
non-thermodynamic transition to a glassy phase in which the energy
(defined as the square of the local cell topological charge) fails to
reach the equilibrium value below a characteristic temperature which
is dependent on the cooling rate. We investigate a correlation
function which exhibits aging behaviour, and follows a master curve in
the stationary regime 
when time is rescaled by a factor of the relaxation time $t_r$. This master curve can be fitted by a von Schweidler law in the late $\beta$-relaxation
regime. The relaxation times can be well-fitted at all temperatures by
an offset Arrhenius law. A power law can be fitted to an intermediate
temperature regime; the exponent of the power law and the von
Schweidler law roughly agree with the relationship predicted by Mode-coupling
Theory.
By defining a suitable response
function, we find that the fluctuation-dissipation ratio is held until
sometime later than the appearance of the plateaux; non-monotonicity
of the response is observed after this ratio is broken, a feature
which has been observed in other models with dynamics involving
activated processes. 
\end{abstract}

\section{Introduction}
Over the last few years, much work has taken place on the topic of the
glass transition and the behaviour of supercooled liquids, both
experimentally and through computer simulations. A great deal of this
activity has involved testing the predictions of Mode-coupling Theory (MCT), which
has proved thus far to reliably describe many key aspects of the
dynamics of glass formers (for a review of these, see \cite{gotzerev}). However, many questions remain,
particularly with regard to the extent to which one might expect a certain
system to satisfy the predictions of MCT.

The bulk of the numerical studies have taken place on binary
Lennard-Jones models, hard sphere systems or amorphous silica, all of which involve a number of
parameters that can be chosen to suit the study e.g. to emulate
experimental systems, or to avoid crystallization etc. By contrast, the
aim of this paper is to investigate a model with as few parameters as
possible, in order to establish  the extent to which the
same features and behaviour can be reproduced. The model we use is
particularly simple in that it is purely topological, involving no
length scale at all. There is also the added advantage that it shows
no tendency to crystallize. Whilst the model is the dual of a
two-dimensional atomic model, the dynamical driving force is not
the same as in that case, and there is no true phase transition.

The lack of spatial co-ordinates within the model invalidates the
investigation of many familiar functions, such as the van Hove
correlation function, the structure factor and the mean squared
displacement. However, suitable alternatives have been identified
which are, in keeping with the ethos of this paper, simpler to calculate.

\section{The Model}
The model we use is that introduced by Aste and Sherrington in
\cite{asteorig}: it is an amorphous
two-dimensional tiling of cells, with 3 edges
incident on a vertex and 2 cells incident on an edge. The average
number of sides of the cells is constrained to be 6 by the Euler
Theorem \cite{euler1, euler2}. These tilings are topologically stable.

If $n_i$ is the number of sides of cell $i$, the topological charge
$q_i=6-n_i$ measures the deviation from the hexagonal configuration
i.e. $q_i$ is a measure of inhomogeneity. Topological charge is
conserved under local rearrangements of the cells.

We define the energy as follows:
\begin{equation}
\label{energy}
E=\sum_{i=1}^N (6-n_i)^2 = \mu_2 N
\end{equation}
where $N$ is the total number of cells in the system. This quantity is
naturally associated with the degree of inhomogeneity in the
configuration: the ground state (a perfect hexagonal tiling) possesses
zero energy. 

\begin{figure}
\begin{center}
\resizebox{!}{120pt}{\includegraphics{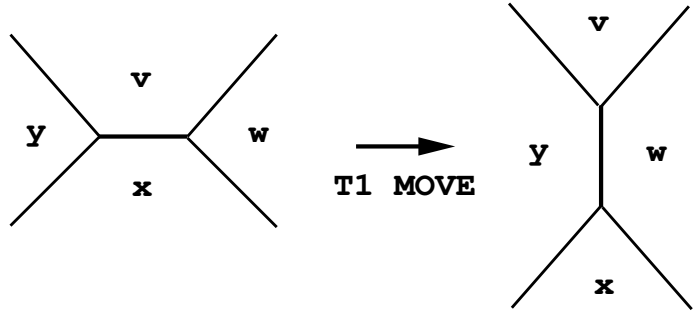}}
\caption{A T1 move: cells $w$ and $y$ gain a side, whilst $v$ and $x$ lose a side. \label{t1}}
\end{center}
\end{figure}

The system is allowed to evolve through moves as illustrated in Figure
\ref{t1}, known in the foam literature as T1 moves. These are local
topological rearrangements exchanging topological charge between four
cells. Two adjacent cells lose a side each and move
apart, and two second-nearest neighbours gain a side each and become
nearest neighbours. The energy change associated with such a move on 4
cells $v,w,x,y$ (as shown in Figure \ref{t1}) with sides $n_v, n_w, n_x$ and $n_y$
is:
\begin{equation}
\label{delta}
\Delta E(n_w,n_y;n_v,n_x)=2(2+n_w + n_y - n_v - n_x)
\end{equation}
We use Glauber-Kawasaki dynamics, which allows evolution of the
system even at zero temperature provided the move decreases the energy
or leaves it unchanged. The probability $P$ of performing a T1 move is
given by:
\begin{equation}
\label{probability}
P(n_w,n_y;n_v,n_x)=\frac{1}{1+\exp\big(\beta \Delta
E(n_w,n_y;n_v,n_x)\big)} (1-\delta_{n_v, 3})(1-\delta_{n_x, 3})(1-\delta_{w,y})
\end{equation}
where $\beta$ is the inverse temperature. The first two $\delta$-functions
forbid the production of two-sided cells, whilst the last $\delta$-function forbids the production of tadpoles (i.e. self-neighbouring
cells); unless these forbidden formations are present at the start, they
will never appear.  

In the following simulations the system consists of $N=9900$ cells, with
periodic boundary conditions. We exhibit results of simulations
starting from configurations obtained by randomly performing $10^4$N T1
moves on a perfect hexagonal tiling. This is equivalent to running the
system at $\beta=0$, and results in an extremely
disordered network, with a value for $\mu_2$ of approximately 13. Time
is measured in units of $N$ attempted moves. 

\section{Relaxation Dynamics}
We study the temporal dependence of the energy of the system by
quenching from a disordered network to $T=2.0$ and allowing the system
to equilibrate at that temperature for $10^4$N attempted
moves. Cooling is then carried out by waiting a time $t=\gamma N$ at
each temperature decrement of $\delta T=0.05$. 
\begin{figure}[t]
\begin{center}
\resizebox{!}{280pt}{\includegraphics{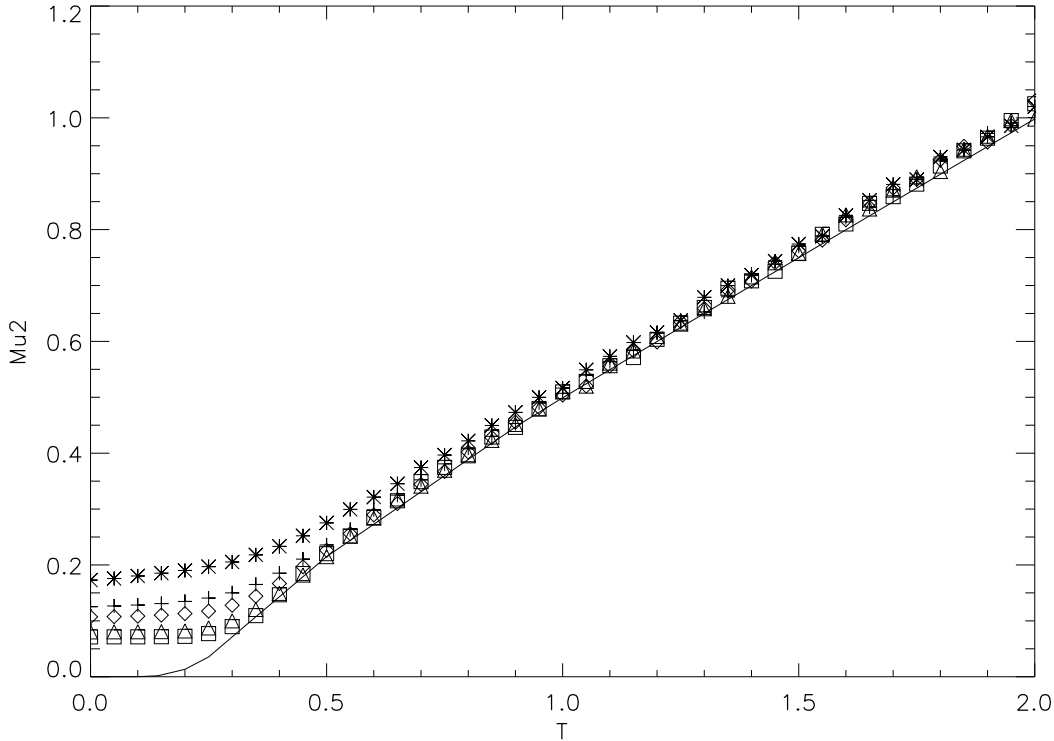}}
\caption{\textbf{The behaviour of $\mu_2$ with temperature.} The
symbols correspond to the following waiting times at each
temperature decrement of $\delta T=0.05$: star -
$10N$, cross - $50N$, diamond - $100N$, triangle - $500N$, square - $1000N$. \label{cool}}
\end{center}
\end{figure}
The results (averaged over 3 runs) are shown in Figure~\ref{cool}. The solid line
superimposed on the graph is the equilibrium curve: the computation of
this is covered thoroughly in \cite{asteorig}. We can clearly see
characteristic glassy behaviour: the system fails to reach equilibrium
even at slow cooling rates for temperatures less than $T \sim 0.2$, and
displays the typical strong dependence of the energy on the cooling
rate, as found at low temperatures in glasses.

We also study the correlations within the system. The choice of
correlation function for this particular system is by no
means obvious. In previous work on this topological model, a
persistence function has been studied, which measures the fraction of
the total cells which have NOT been involved in a T1 move
\cite{asteorig}. However, this provides only limited information about the system: by the very
definition of the function, information about a cell is 
thrown away once it has been involved in a move. If the system was in a
metastable state in which it performed then undid many T1 moves, this
would not be revealed.

Instead we have have chosen to use self-correlation functions: the
first of these is
\begin{equation}
\label{correl}
C(\tau,\tau + t)=\frac{\sum_{i=1}^N \big(n_i(\tau)-6\big)\big(n_i(\tau + t)-6\big)}{\sum_{i=1}^N \big(n_i(\tau)-6\big)^2}
\end{equation}
We add in passing that the results for this particular function differ
negligibly in a qualitative sense from those for an energy-energy self-correlation function
such as that used in conjunction with the backgammon model
\cite{backgammon1, backgammon2}. 

\begin{figure}
\subfigure[$C(t)$ for $\beta=1.0, 2.0, 2.5, 3.0, 3.5, 4.0$ (from left to
right). These are equilibrium results.]{\label{eqcorrel2}\resizebox{!}{178pt}{\includegraphics{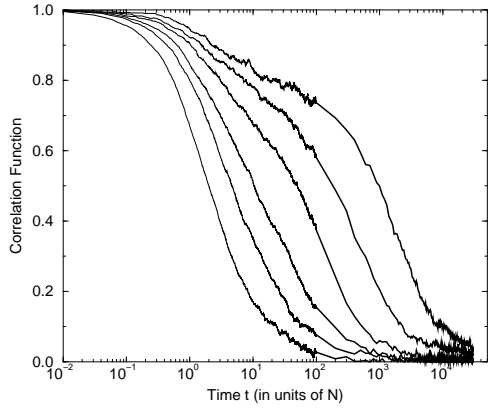}}}
\subfigure[$C(\tau,\tau+t)$ for (from left to right) $\beta=$$4.5$, $5.0$, $ 6.0$, $7.0$. 
$\tau=60,000N$. These are non-equilibrium results.]
{\label{highbeta}\resizebox{!}{178pt}{\includegraphics{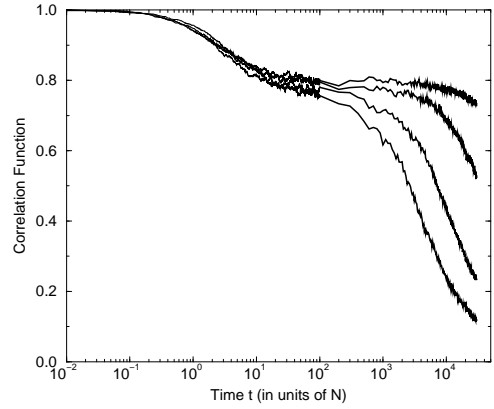}}}
\subfigure[$C(\tau,\tau+t)$ for $\beta=4.0$. From left to right,
the curves correspond to $\tau=10^2N, 10^3N$ and $10^4N$.]{\label{4beta}\resizebox{!}{174pt}{\includegraphics{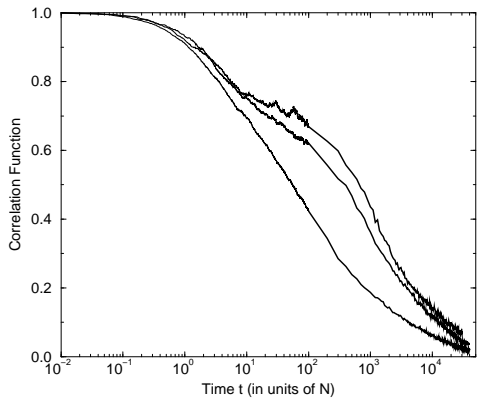}}}
\subfigure[$C(\tau,\tau+t)$ for $\beta=6.0$. From left to right,
the curves correspond to $\tau=10^2N, 10^3N$ and $10^4N$.]{\label{6beta}\resizebox{!}{180pt}{\includegraphics{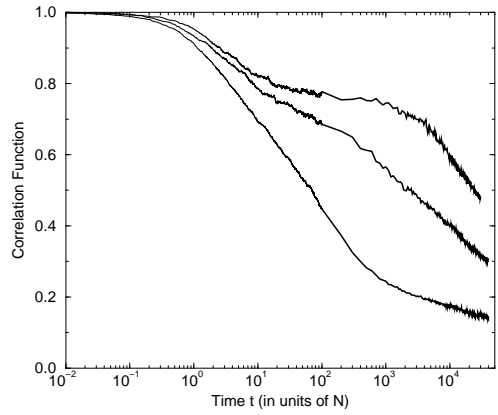}}}
\caption{Correlation functions both in and out of equilibrium (each averaged over
3 runs).\label{eq}}
\end{figure}

We shall first present the results in equilibrium, where the
correlation functions are stationary and possess no $\tau$
dependence. Thus the correlator $C(\tau, \tau + t)$ becomes $C(t)$, a function
of $t$ only. In order to obtain stationarity, the system was allowed to
equilibrate for $60,000N$ at each temperature before measurements were
taken, and for $200,000N$ at $\beta=4$. The energy was found to have equilibrated in each case and no
$\tau$ dependence was found. For $\beta>4$ it became impossible
to achieve equilibrium within any practical timescale; even if the
curves appeared stationary, the energy had not equilibrated.

Figure \ref{eqcorrel2} shows the correlation functions for
a range of temperatures. At low values of $\beta$ (high temperatures)
the correlator decays directly to zero. As the value of $\beta$ is
increased, we see a shoulder develops, indicating that two-step relaxation is
taking place. If we were able to attain stationarity at higher values
of beta we would see this shoulder broaden to a plateau. The plateau can be
seen more clearly in Figure \ref{highbeta}, which shows the correlation function for higher values
of $\beta$, with the waiting time of $60,000N$;
however, one must keep in mind that this figure shows non-equilibrium results.
Figures \ref{4beta} and \ref{6beta} show the strong $\tau$ dependence
that exists out of equilibrium.

We can study the correlation function within the stationary
regime. The relaxation around the shoulder or plateau, and the early
stages of the departure from it, are known as the
$\beta$-relaxation regime. Mode-coupling Theory predicts the
behaviour of conventional correlation functions in the late part of this regime
(i.e. during the plateau and the departure from it) to be governed
by a von Schweidler law. This states that in stationary conditions the
correlator $C(t)$ can be described as follows:
\begin{equation}
\label{vonsch}
C(t)=f-B \hspace{1pt} \big(t/t_{r}(T)\big)^b
\end{equation}
where $f$ is the plateau height (also known as the non-ergodicity parameter),
and both $B$ and $b$ are positive constants. All three should be
temperature-independent, and in addition $b$ should be
independent of the choice of correlator. However, this theory is based on
atoms interacting via a two-body potential and studies conventional
correlation functions with explicit position dependence, such as the
density-density correlator (for a recent review see \cite{gotzerev}). Thus it
is interesting to plot our rather unusual 
correlation function against rescaled time $t/t_r$ to investigate the
predictions of MCT in our case.

\begin{figure}[t]
\begin{center}
\resizebox{!}{178pt}{\includegraphics{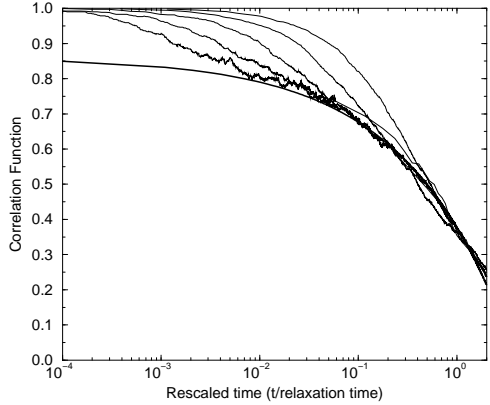}}
\caption{$C(t)$ for $\beta=1.0, 2.0, 2.5, 3.0, 3.5, 4.0$  (from right to
left) against rescaled time. The solid curve is a von
Schweidler fit as in equation \ref{vonsch} with $f=0.86$, $B=0.483$ and $b=0.42$.\label{rescale3}}
\end{center}
\end{figure}

Following the procedure of Kob \cite{kobreview}, we define the relaxation time
$t_{r}$ in Equation \ref{vonsch} as the time at which $C(t)$ first drops below
$\mathrm{e}^{-1}$.  Figure \ref{rescale3} shows the correlation functions for a range of beta against
rescaled time. All the functions collapse onto a master curve in the
region after the plateau or shoulder, thus
obeying the time-temperature superposition principle of MCT. In the
late $\beta$-relaxation regime this curve can be fitted by a  von
Schweidler law with $f=0.86$, $B=0.483$ and $b=0.42$ as shown on Figure~
\ref{rescale3}.  The data fits well to this law, although because of
the noise the margin of error on the fitting parameters is large.

\begin{figure}
\begin{center}
\subfigure[\textbf{The logarithm of $t_r$ against inverse
temperature.} The solid curve is an offset Arrhenius fit
i.e. $t_r =
0.0195 \exp(2.85\beta) + 3.13$]{\label{rel1}\resizebox{!}{220pt}{\includegraphics{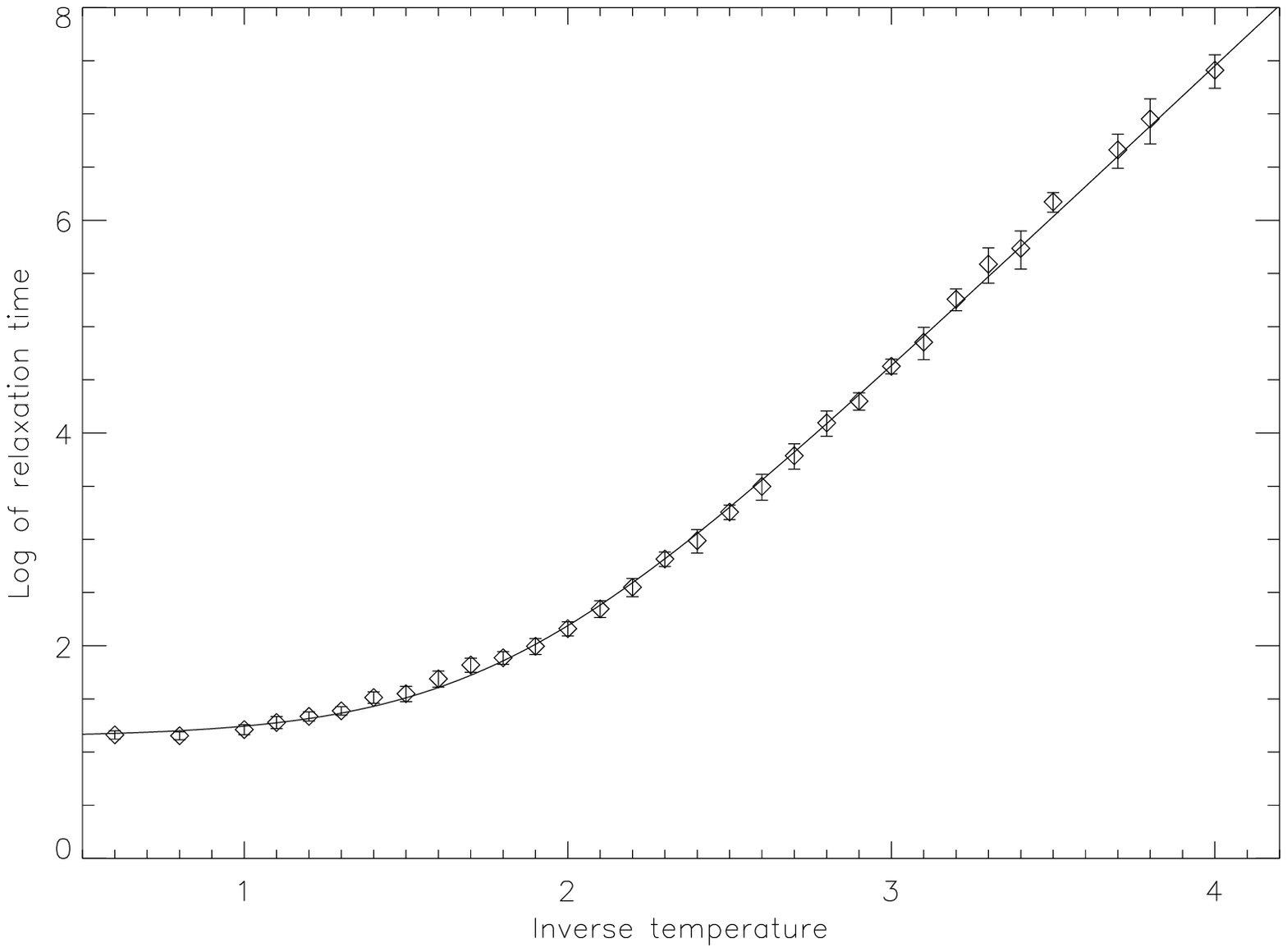}}}
\end{center}
\subfigure[\textbf{The logarithm of $t_r$ against $(T-T_o)^{-1}$.} The
straight line is a best-fit Vogel-Fulcher law with
$T_o=0.202$.]{\label{rel2}\resizebox{!}{150pt}{\includegraphics{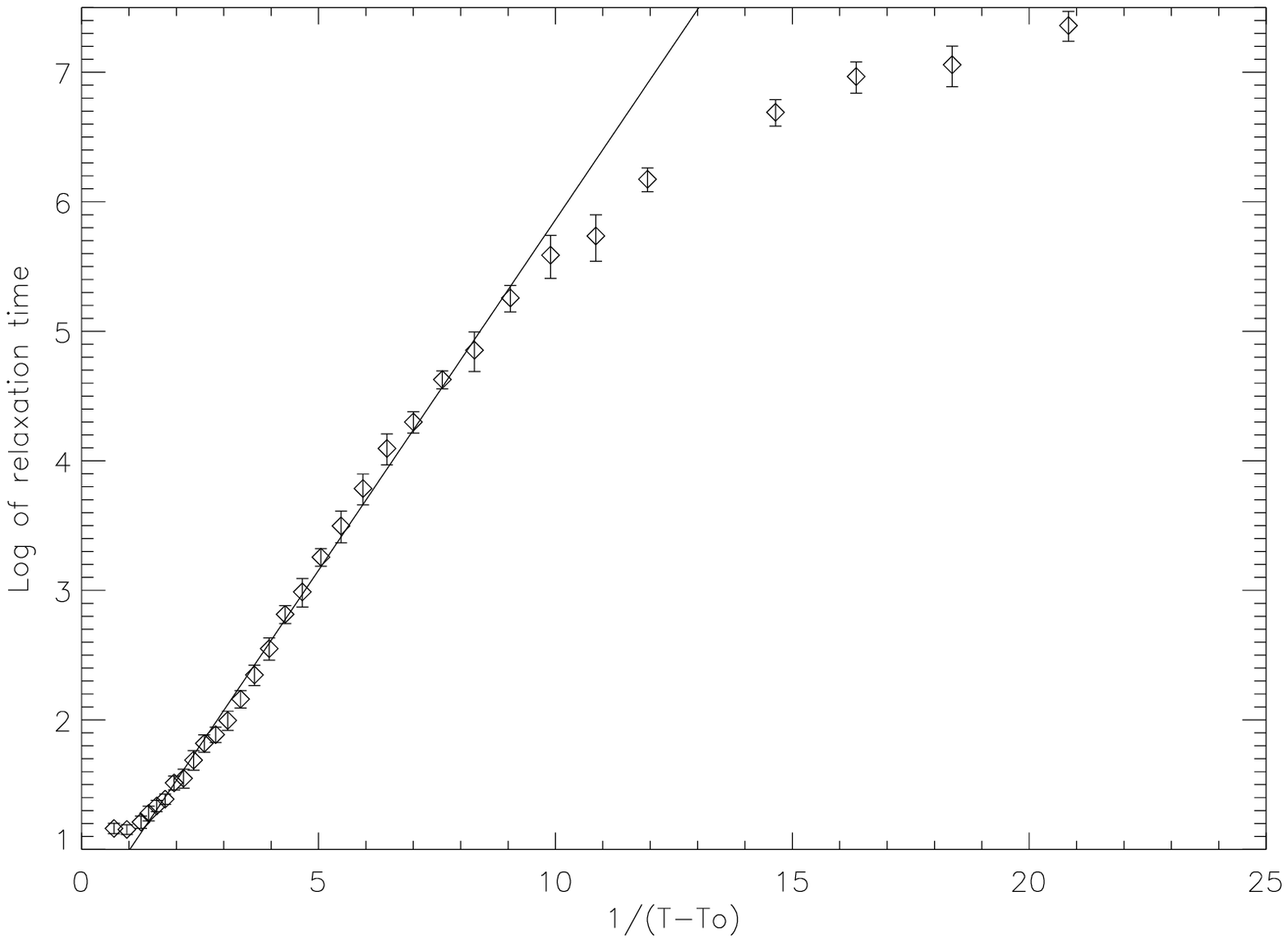}}}
\subfigure[\textbf{The logarithm of the inverse relaxation time
against the logarithm of $(T-T_c)$.}
The straight line is a best-fit power law with
$T_c=0.218$.]{\label{rel3}\resizebox{!}{150pt}{\includegraphics{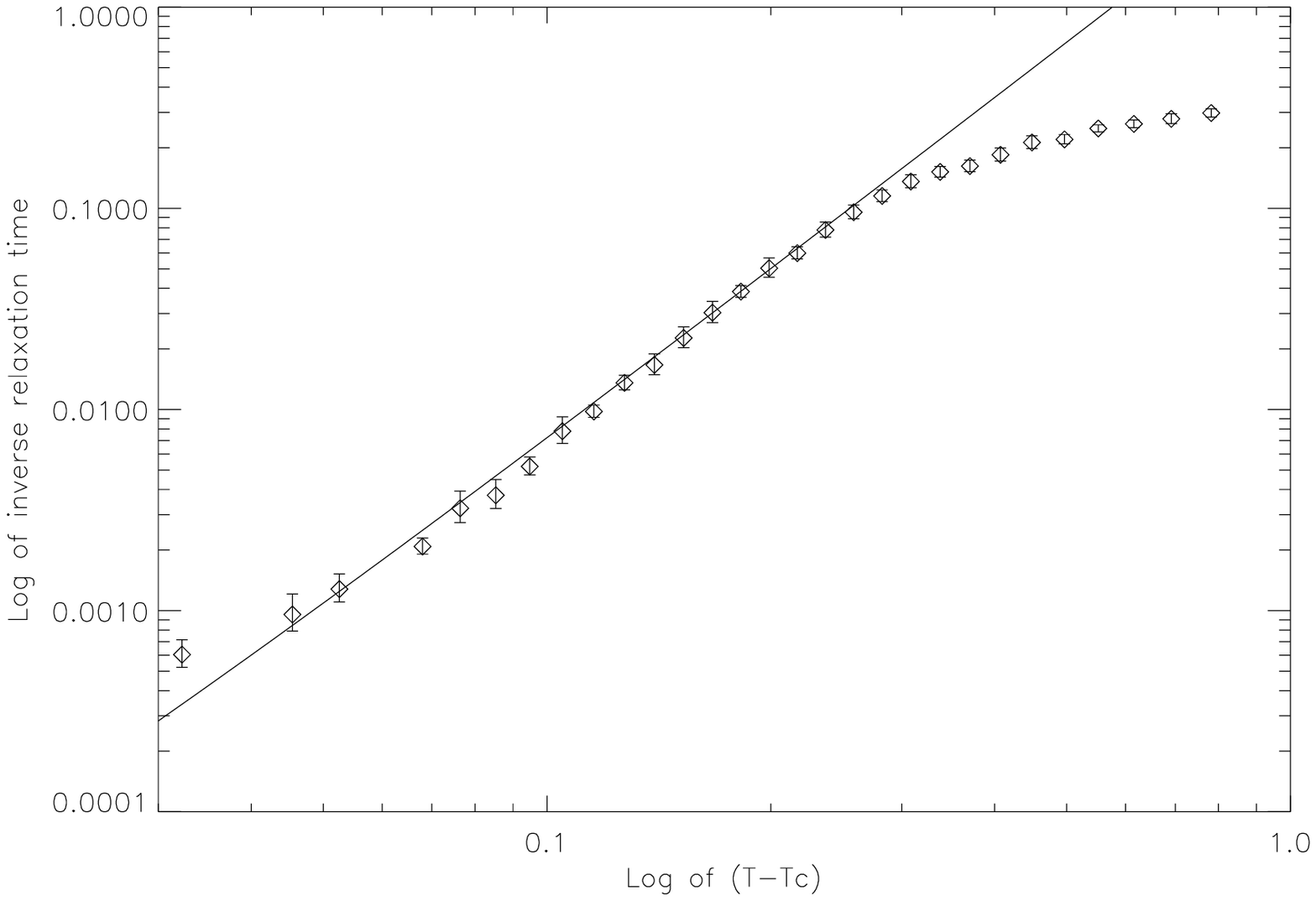}}}
\caption{\textbf{The behaviour of the relaxation time with temperature.}}
\end{figure}

Figure \ref{rel1} shows the relaxation time against inverse
temperature. This can be fitted extremely well by the following
offset Arrhenius function: 
\begin{equation}
t_r = A+B \hspace{1pt} \mathrm{e}^{C/T}
\end{equation}
where $A, B, C$ are constants. This
is plotted on Figure \ref{rel1}, with $A=3.13, B=0.0195$ and
$C=2.85$. Thus at very low temperatures (high inverse temperature) the
system is displaying Arrhenius behaviour, which is characteristic of
strong glass formers. 

In the high temperature regime, one can fit the Vogel-Fulcher law
typical of fragile glass-formers i.e. $t_r~=~A~\mathrm{e}^{C/(T~-~T_o)}$
with $T_o=0.202$ (see Figure \ref{rel2}), but it does not suit the
data as well as the offset Arrhenius curve. 

The relaxation time is also subject to predictions by MCT, which
states it should behave as:
\begin{equation}
\label{reltime}
t_r \propto (T-T_c^{MCT})^{-\gamma}
\end{equation}
close to a critical Mode-coupling temperature $T_c^{MCT}$, with $T_c^{MCT}$ defined through
this equation. By performing such a fit on our data, we find that
$T_c^{MCT}=0.218$, as shown on Figure \ref{rel3}, with $\gamma=2.86$. Neither this law nor the
Vogel-Fulcher law fit the data well very close to $T_c^{MCT}$, although they
both fit reasonably well slightly above $T_c^{MCT}$. However, it would seem
that whilst one can fit both of these functions in certain regions, the best description
at all temperatures is provided by the offset Arrhenius law.

Mode-coupling Theory predicts a relationship between the exponent
$\gamma$ in Equation~\ref{reltime} and the exponent $b$ in
Equation~\ref{vonsch}. They are linked by:
\begin{equation}
\label{gamma}
\gamma=\frac{1}{2a} + \frac{1}{2b}
\end{equation}
where:
\begin{equation}
\label{Gamma}
\frac{\Gamma(1+b)^2}{\Gamma(1+2b)}=\frac{\Gamma(1-a)^2}{\Gamma(1-2a)} 
\end{equation}
Using our value of $b=0.42$, we find a value for $\gamma$ of
$3.12$, which can be compared with the power law fit of $\gamma=2.86$. These two results agree to
within $10\%$, which is reasonable given that the  noise of the data makes it is very difficult to fit
either exponent accurately.

\section{Response Functions}
A clear indication of the presence of aging is the violation of the
fluctuation-dissipation theorem \cite{bouchcugreview, parisifdr}, which
we shall briefly review. In
equilibrium the response $R_A$ of an observable A to an applied
conjugate field relates to the appropriate time-correlation function
$C_A=<A(t)A(0)>$ as follows (in units of $k_B=1$):
\begin{equation}
\label{eqresp}
R_A(t)=-T^{-1} \frac{\partial C_A(t)}{\partial t}
\end{equation}
Out of equilibrium the relationship must be generalised since the
correlation function is now a two-time quantity:
\begin{equation}
\label{noneqresp}
R_A(\tau, t + \tau) = -T^{-1} X_A (\tau, t+ \tau) \frac{\partial
C_A(\tau, t+ \tau)}{\partial \tau}
\end{equation}
with $X_A(\tau, t + \tau)$ defined by the above equation. If $X_A(\tau, t
+ \tau)$ is equal to 1, equation \ref{noneqresp} reverts to the
equilibrium case. 

In order to investigate autocorrelations, one can apply a field at time $\tau$ which has a magnitude
$h_o$, but is randomly positive or negative across the entire
system. That field is left switched on, and the integrated response
$G(\tau, t+\tau)$ is measured, where:
\begin{equation}
\label{intresp}
G(\tau, t+\tau)=h_o \int_{\tau}^{t+\tau} R_A(t', t+ \tau) dt'
\end{equation}
For many systems it has been observed that for $t$ and $\tau$ both
large, $X_A(\tau, t + \tau)$ depends on $t,\tau$ only through the
correlator i.e. $X_A(\tau, t + \tau)=x\big(C(\tau, t + \tau)\big)$. If
that is indeed the case, one finds:
\begin{equation}
\label{intx}
\frac{-T G(\tau, t+\tau)}{h_o} = \int_{1}^{C} x(C')dC'
\end{equation}
Under those conditions the slope of a parametric plot of $-TG(\tau, t+\tau)/h_o$
against the relevant correlation function is the interesting quantity,
since:
\begin{equation}
\label{Xform}
\frac{\partial}{\partial C} \Big(\frac{-T G(\tau, t+\tau)}{h_o}\Big)=  x(C)
\end{equation}
One would expect to find a slope of
exactly $-1$ where the fluctuation-dissipation ratio is upheld. Where the
ratio is broken, the slope gives information about the form of $X_A(\tau, t + \tau)$.

In our particular case, we apply a perturbation $h_o \sum_{i=1}^N \epsilon_i
(n_i - 6)$ at time $\tau$. $\epsilon_i$ is randomly assigned to be $+1$
or $-1$, and $h_o$ is chosen carefully to ensure linear response
whilst also obtaining a reasonable signal-to-noise ratio. We follow the integrated response $G(\tau,t+\tau)=\sum_{i=1}^N
\epsilon_i (n_i-6)$. It is easy to see that the
appropriate correlation function for this observable is of the form of
that given in equation \ref{correl}, although one must be careful with
normalisation factors. The data is in general very noisy and it is
necessary to average over at least 8-10 different field configurations
to be able to reach any conclusions.

Before presenting results, we draw attention to the unusual nature of
our integrated response function. At low temperatures almost all the
cells within the system are 5, 6 or 7-sided, with the vast majority
being 6-sided. This means that a large proportion of the system
\textit{makes no contribution to the response
whatsoever}. Furthermore, processes
which reduce the energy by turning pentagons and heptagons into
hexagons in fact reduce the response. This will be discussed in more
detail in the following paragraphs; in the meantime, it is sufficient
to remember that the integrated response function used here has some
unusual features.
\begin{figure}
\subfigure[$\beta=4$, $\tau=10^4N$ (upper curve) and $10^3N$ (lower
curve)]{\label{4resp}\resizebox{!}{150pt}{\includegraphics{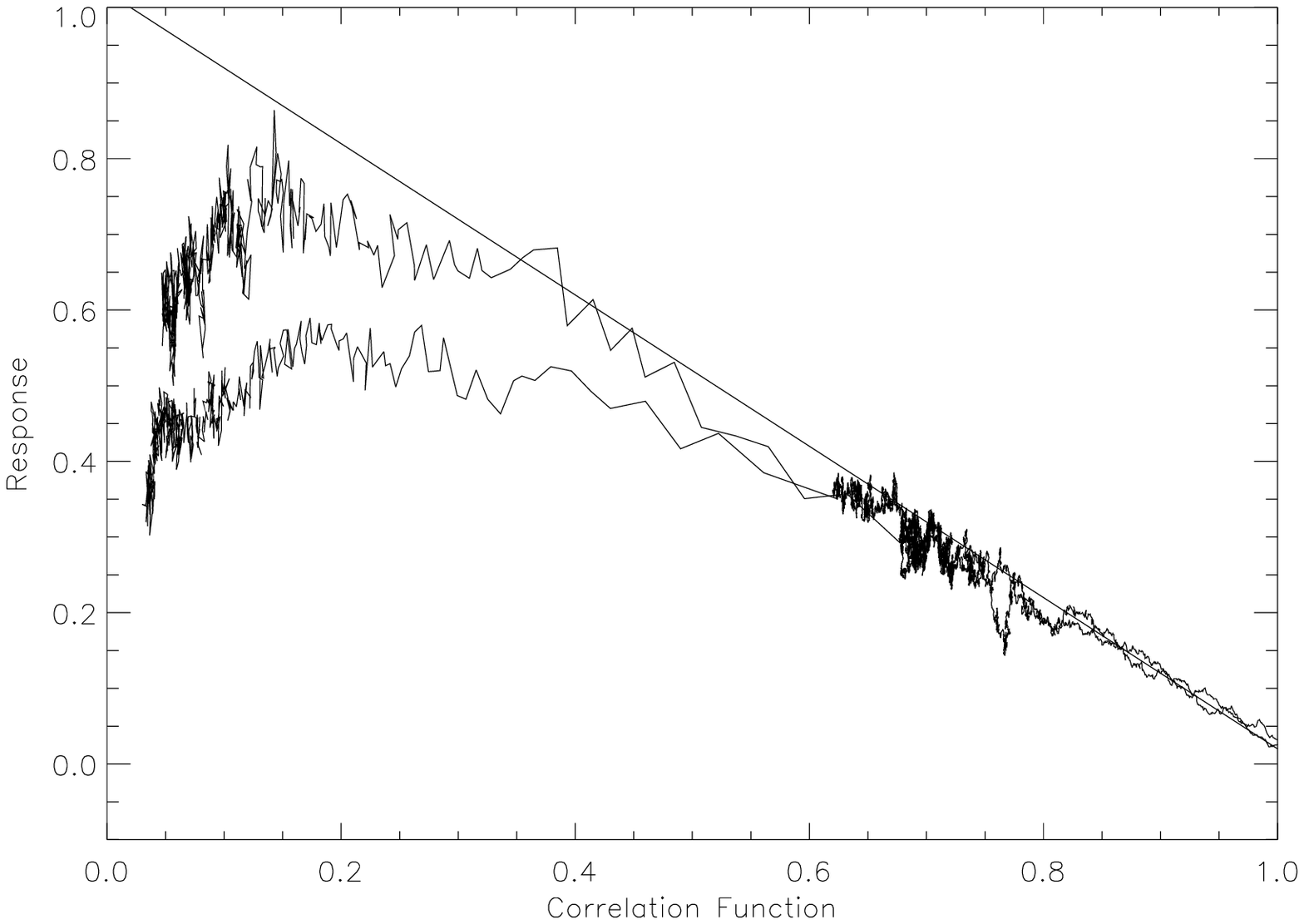}}}
\hfill
\subfigure[$\beta=5$, $\tau=10^4N$ (upper curve) and $10^3N$ (lower curve)]{\label{5resp}\resizebox{!}{150pt}{\includegraphics{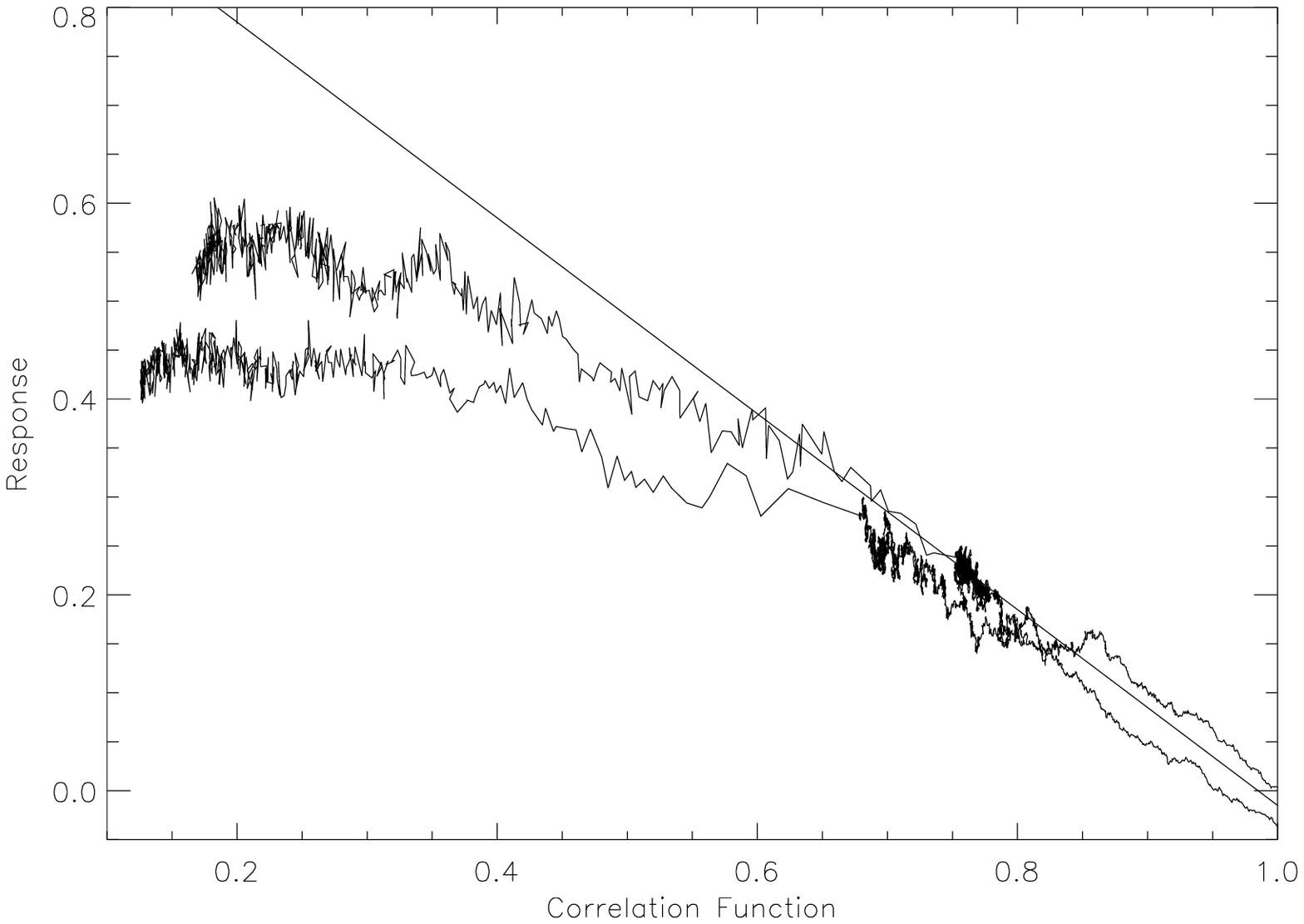}}}
\caption{$\frac{-TG(\tau,\tau + t)}{h_o}$ plotted parametrically
against $C(\tau,\tau + t)$ at different $\beta$ and $\tau$ (averaged over
10 runs).\label{resp}}
\end{figure}

Figure \ref{resp} shows the results for $\beta=4$ and $5$. These are
non-equilibrium results in all cases, as the waiting times are not
long enough for the system to have equilibrated. The
superimposed straight lines have slope -1, but are not best fits; they
are there to give the eye a comparative reference. These plots should
be examined from right to left i.e. short times are on the right, where
$C(\tau, \tau + t)$ is close to $1$, and long times are on the
left. We see that for short times the
fluctuation-dissipation ratio is upheld; the slope is -1 for all
values of $\tau$, although the intercept does of course vary. In fact, the ratio is not
broken until $C(\tau, \tau + t)$ has decreased to $\sim 0.6$: this is
considerably lower than the value of $C(\tau, \tau +
t)$ at which the plateau occurs. Thus the fluctuation-dissipation ratio is obeyed for a
timescale larger than that of the onset of the plateau.  After the
ratio has been broken, curves for different $\tau$ behave differently,
tending towards the equilibrium straight line of slope $-1$ as $\tau$
increases. The non-monotonicity displayed clearly in Figure
\ref{4resp} occurs in every case, although at lower temperatures one
cannot run long enough to see it, as in Figure \ref{5resp}.
Non-monotonicity of the response has been
noted in several other models and will be discussed briefly in the
next section. In the meantime, with regards to this model this unusual
behaviour can be understood if we now turn to a
discussion of the processes dominating the evolution of the system.

\begin{figure}[h]
\subfigure[Zero energy T1
moves.]{\label{zerot1}\resizebox{!}{180pt}{\includegraphics{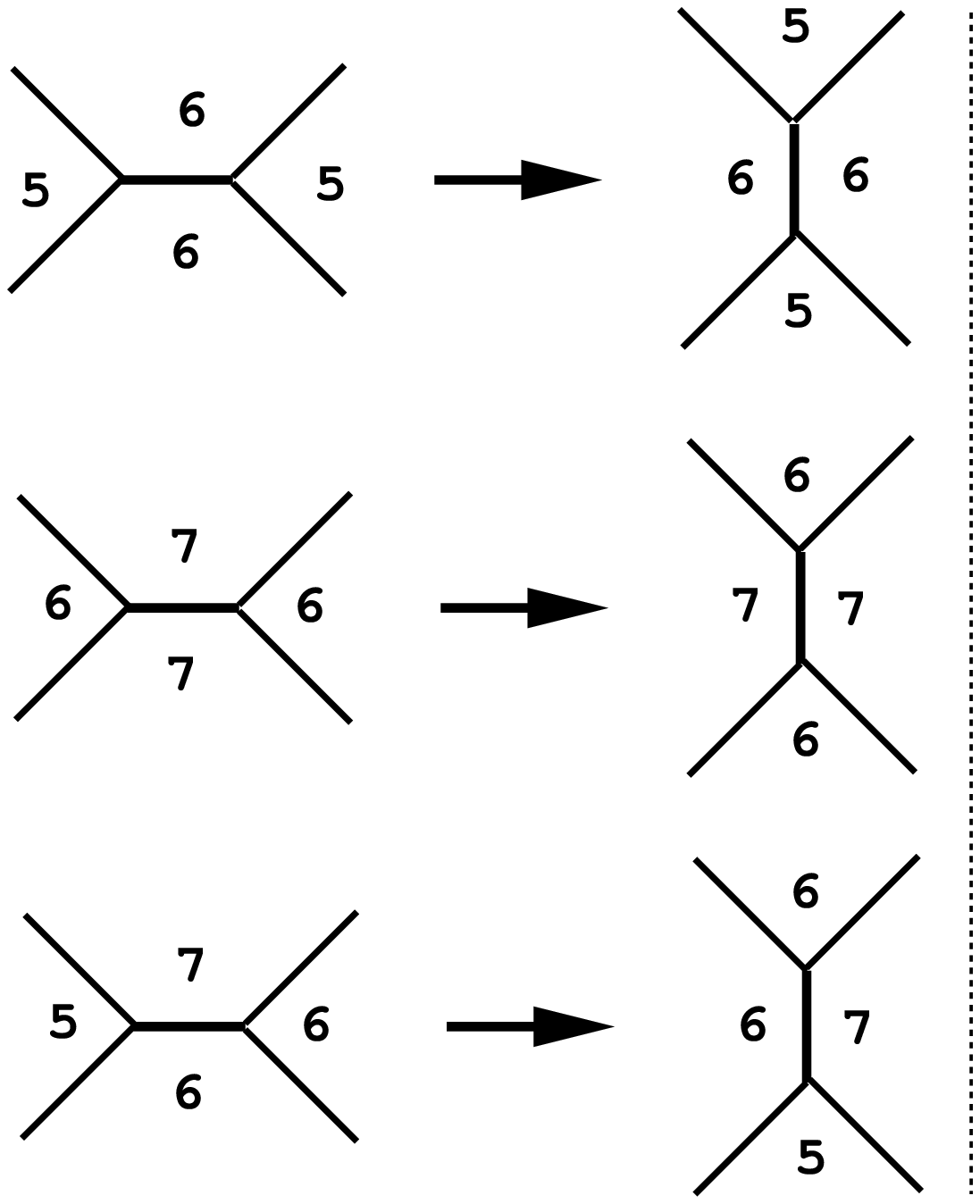}}}
\hfill
\subfigure[T1 moves which reduce the energy.]{\label{neg}\resizebox{!}{180pt}{\includegraphics{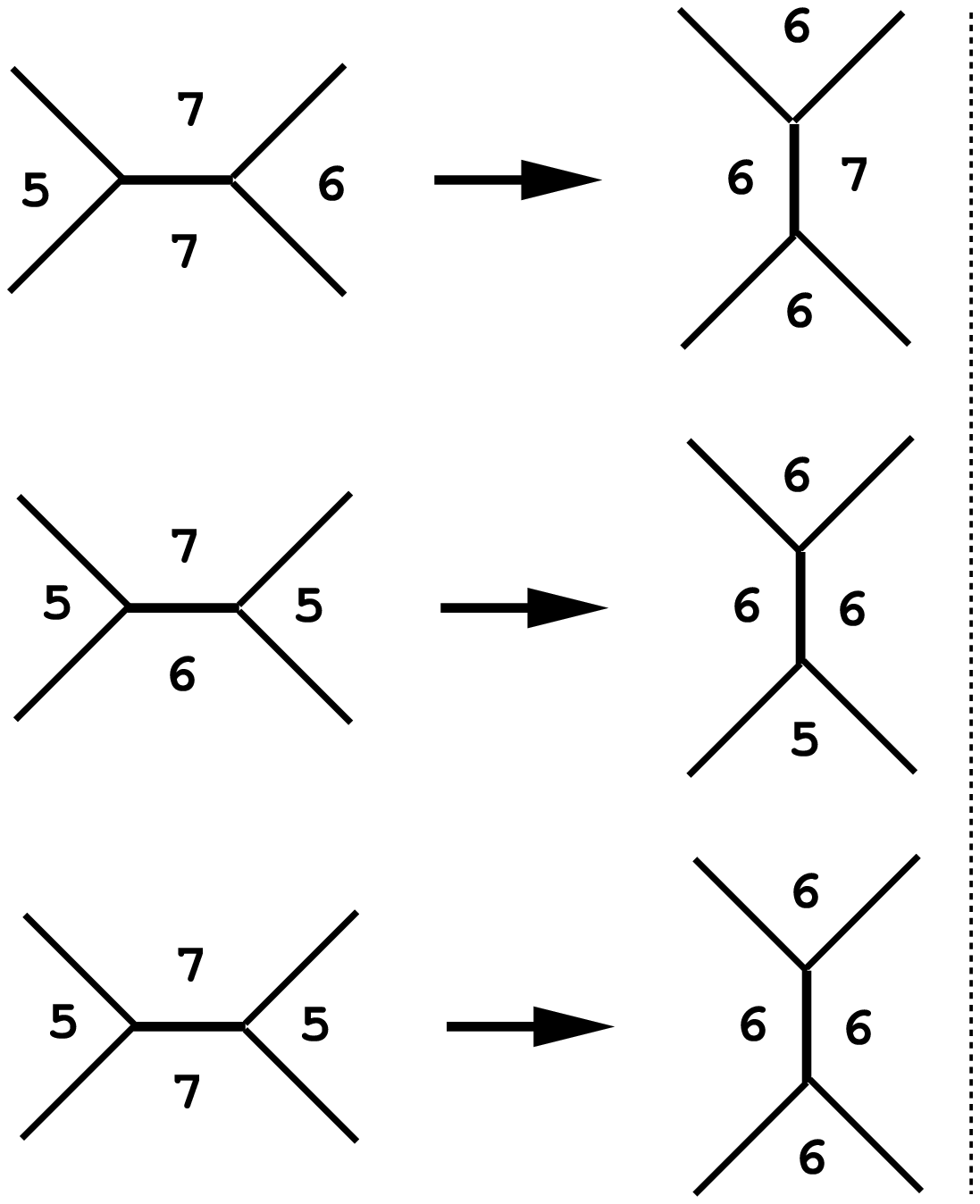}}}
\hfill
\subfigure[Unlikely zero energy T1
moves.]{\label{unlikely}\resizebox{!}{110pt}{\includegraphics{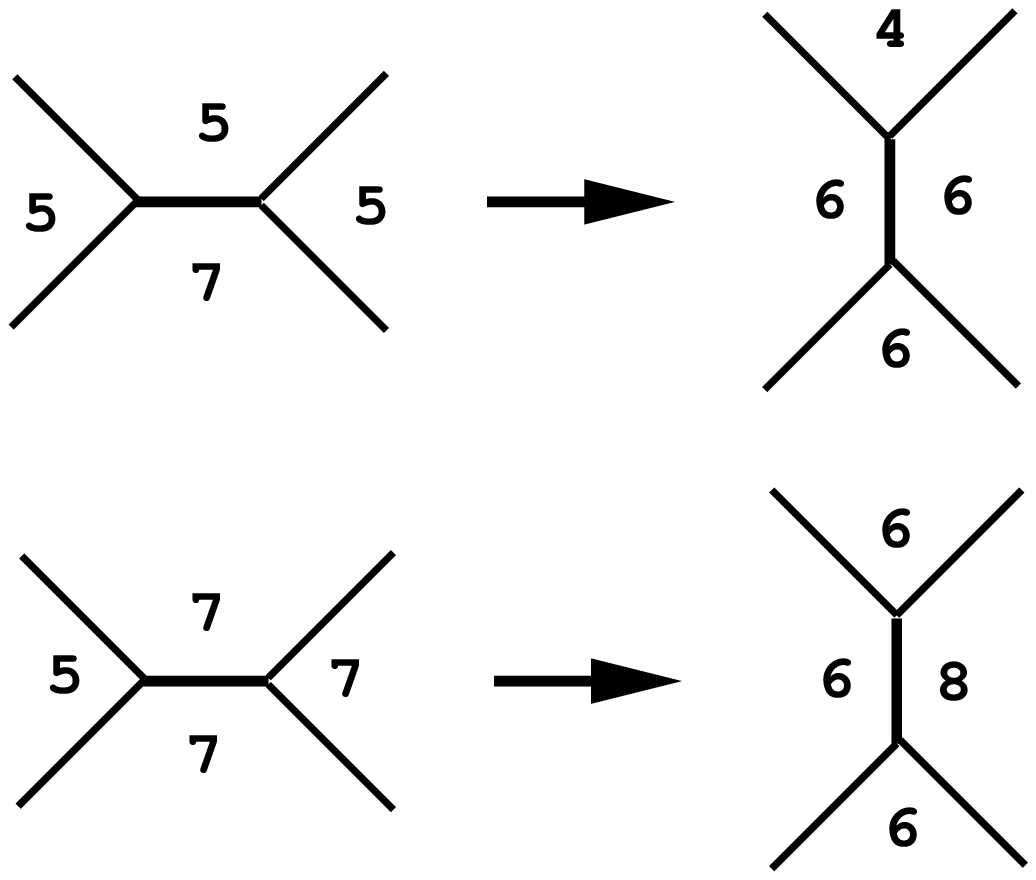}}}
\caption{The possible T1 moves at low temperatures.\label{moves}}
\end{figure}

At low temperatures, out of equilibrium, moves which increase the
energy occur very rarely; we can consider the system to evolve by a
combination of zero-energy moves and moves which reduce the
energy. After a short time ($\sim 10^3N$) the system consists almost
entirely of 5, 6 and 7-sided cells. Under these conditions the
possible moves are as shown in Figure \ref{zerot1} and \ref{neg}. The moves in
Figure \ref{unlikely} are
considered very unlikely due to the low probability of finding four
non-hexagons clustered together; this is corroborated by the extreme
rarity of finding octagons and rectangles present. We see from Figure \ref{zerot1}
that in certain topological arrangements, pentagons and heptagons can
effectively diffuse freely through the system; in particular,
pentagons and heptagons can `pair up' to move through the system
together with no energy cost. To reduce the energy,
one must annihilate two or more `defects' (i.e. non-hexagons) as in
Figure \ref{neg}; this requires the absorption of a 5-7 pair by an
isolated pentagon or hexagon, or the annihilation of two 5-7
pairs. This brings us to a conceptual picture containing both fast and slow
dynamics: the fast dynamics is due to the rapid diffusion of 5-7 pairs
moving freely through the system, whereas the slow dynamics is due to
annihilation processes. The latter can be broken down into two
different types: absorption of a 5-7 pair by either a heptagon or a
pentagon (which can also rearrange the network such that
isolated pentagons and heptagons become 5-7 pairs), and the complete
annihilation of two 5-7 pairs. Of course, in equilibrium an equal number of 5-7 pairs are
created as annihilated/absorbed, but in the relaxation to equilibrium,
annihilation has the upper hand as the system starts from a configuration in
which there are more 5-7 pairs present.

The effect of the field is to try and `pin' defects onto the
appropriate cell i.e pentagons try to settle on sites with $\epsilon_i
=+1$ and heptagons on $\epsilon_i=-1$. Thus with the field switched on, the
diffusion processes move the defects around until they are on the
appropriate sites, and in this way increase the value of the response
$-TG(\tau,\tau + t)/h_o$. However, annihilation processes
remove defects from the system entirely, thus reducing the response
and also reducing $C(\tau, \tau +t)$. We have competition between the
annihilation processes and the diffusive processes; when annihilation
dominates, the slope on the parametric plot
becomes positive, as seen for small values of $C(\tau, \tau +t)$ in Figure \ref{4resp}. If one looks instead at a quantity such as
$-TG(\tau,\tau + t)/h_o \mu_2$ (in some sense a `response
per defective cell') one finds that the slope remains negative: this
tells us that the diffusion processes are succeeding in placing more of 
the defects that remain on the appropriate sites, but the number of
those defects present in the system is declining. The value of the
response reached when the system finally achieves equilibrium depends
on the waiting time $\tau$, and increases as $\tau$ increases; this is
because both the response and the correlation
function have been normalised by a factor that is equivalent to the
energy at time $\tau$ (see equation \ref{correl}).

This picture of diffusion and annihilation/creation processes is
helpful to an understanding of the plateaux in the correlation
functions at low temperatures. The initial descent to the plateau is
due to diffusing 5-7 pairs, which quickly move the network away from
the starting configuration. However, the lone defects are trapped, and
can only be freed by a move which costs energy or by absorption of a 5-7 pair; both
of these processes occur on a timescale which is
temperature-dependent, and thus the length of the plateau itself is
dependent on the temperature. This picture of
trapped or caged defects is conceptually similar to that typically
used when dealing with Lennard-Jones binary models~\cite{kobreview}\cite{kobgleim}.
 
\section{Concluding remarks}
In summary, we have studied a simple two-dimensional topological
glassy model, testing some of the predictions of Mode-coupling theory
for this system. We find that this theory can indeed be used to describe
the behaviour of the system in so far as we have investigated,
although this description fits some features more closely than others.
The correlation function follow a von Schweidler law as predicted in
the late $\beta$-relaxation regime, and the exponent of this roughly agrees 
with the exponent of the power law fitted to the relaxation time
data. However, the relaxation times fit better to an offset Arrhenius
law in all temperature regimes.

Investigation of the fluctuation-dissipation ratio reveals an
interesting feature, namely non-monotonicity of the response. This is
a feature which has also been observed in other simple models; for
example, vibrated granular media \cite{granular}, constrained Ising
chains and the
Backgammon model in one dimension \cite{crisanti}, and also recently in a
two-dimensional short-range spin model with uniform ferromagnetic
3-body interactions \cite{juanpe}. In each case, the dynamics at low
temperatures can be considered to involve activation over energy
barriers; as mentioned in \cite{juanpe}, this raises the question as
to whether this is a generic feature of models with activated
dynamics.

\section{Acknowledgements}
The authors would like to thank T. Aste and J. P. Garrahan for helpful
discussions, and \hbox{EPSRC(UK)} for financial support - DS for
research grant GR/M04426, and LD for research studentship 98311155.

\end{document}